\documentclass[epj]{svjour}
\usepackage{graphics}
\usepackage{amssymb}

\begin{document}
\title{Optimal disorder for segregation in annealed small worlds}
\author{Santiago Gil and Dami\'an H. Zanette
\thanks{Also at Consejo Nacional de Investigaciones
Cient\'{\i}ficas y T\'ecnicas, Argentina.}} \institute{Centro
At\'omico Bariloche and Instituto Balseiro, 8400 Bariloche,
R\'{\i}o Negro, Argentina}
\date{Received: date / Revised version: date}

\abstract{We study a model for microscopic segregation in a
homogeneous system of particles moving on a one-dimensional
lattice. Particles tend to separate from each other, and evolution
ceases when at least one empty site is found between any two
particles. Motion is a mixture of diffusion to nearest-neighbour
sites and long-range jumps, known as annealed small-world
propagation. The long-range jump probability plays the role of
the small-world disorder. We show that there is an optimal value
of this probability, for which the segregation process is
fastest. Moreover, above a critical probability, the time needed
to reach a fully segregated state diverges for asymptotically
large systems. These special values of the long-range jump
probability depend crucially on the particle density. Our system
is a novel example of the rare dynamical processes with critical
behaviour at a finite value of the small-world disorder.
} \maketitle

Segregation --the spatial separation of different elements in a
multi-component system-- is a paradigmatic form of complex
behaviour, present in a wide variety of natural phenomena. It can
occur both at the microscopic level, driven by repulsive
interactions between individual components, and at the macroscopic
level, as in the case of phase separation due to collective
processes. In many instances, it leads to the spontaneous
emergence of spatio-temporal order, in the form of coherent
structures with non-trivial evolution.

In physical systems, spatial segregation is conspicuous in the
vicinity of phase transitions, as the result of the amplification of
microscopic fluctuations \cite{trans}. An effect of similar origin
is observed in some elementary chemical reactions limited by
diffusion. It is well known, for example, that the binary
annihilation of two diffusing chemicals $A$ and $B$, $A+B \to 0$,
leads to the formation of single-species spatial domains just after
the very first stages of the process, starting from a uniform
mixture of the two species \cite{anih}.

In the realm of biology, segregation at the molecular and cellular
level plays a prominent role in morphogenesis, i.e. in the
differentiation of the various tissues, organs, and functional parts
of living organisms \cite{morphog}. At the level of ecosystems,
environmental heterogeneities may lead to population segregation,
with the appearance of specific localized niches, when the relative
fitness of competing species varies in space \cite{evol}. In human
populations, the combination of cultural diversity with xenophobic
feelings is able to disrupt society, generating hatred and
aggression where cooperative behaviour should otherwise emerge
\cite{cult}.

Homogeneous ecological and social systems, with no substantial
differences between their individuals, often exhibit a different
form of segregation, not necessarily associated with the formation
of spatial structures. Single individuals --or small groups of
closely related individuals, such as families or clans-- may
choose to separate from each other as much as possible, all over
the available spatial domain. This spreading is related to a more
efficient exploitation of the resources associated with land, and
has certainly been crucial to the geographical dissemination of
biological species over the ages \cite{evol}.

In this paper, we explore a physically-inspired model for this
kind of segregation. We pay particular attention to the dynamical
process that leads, from a random distribution of the population,
to a stationary state where individuals have mutually separated
such that no contiguous neighbours are found. Individuals are
allowed to move diffusively and, occasionally, they can perform
long jumps to randomly selected sites. This form of transport,
inspired in the connection structure of small-world networks, is
discussed in the next section. In Sect. \ref{num1}, we present our
first numerical results, and show that the time needed to reach
full segregation depends non-monotonically on the probability of
long jumps. Remarkably, in fact, the time is minimal for an
intermediate value of that probability. Section \ref{anal} is
devoted to develop an analytical approximation of our problem,
which makes it possible to explain some of the numerical results.
Further simulations are presented in Sect. \ref{num2}, where we
validate predictions of the analytical approximation and complete
the study of our system. Results are summarized and discussed in
the last section.

\section{Segregation in annealed small worlds}

We consider an ensemble of $N$ particles distributed on a
one-dimensional array of $L$ sites ($L\ge N$) with periodic
boundary conditions. The density is $\rho=N/L$. An exclusion
principle holds so that, at any time, each site may be occupied by
at most one particle. Particles tend to segregate, and a particle
is said to be {\it frustrated} if at least one of its two
nearest-neighbour sites is occupied. Frustrated particles are
allowed to move, in order to reach a non-frustrated state. Motion
of frustrated particles is governed by annealed small-world
transport, as described in the following.

Small-world networks have been defined to capture two essential
properties of real social structures, namely, the small average
distance between any two individuals (small-world effect), and
their high clustering \cite{wattstr}. They have been used, as toy
models, to demonstrate the role of the underlying interaction
pattern in dynamical processes of socio-economical and ecological
inspiration \cite{sw1,sw2,sw3}. These networks are built starting
from an ordered lattice of relatively high connectivity. A
prescribed fraction of links is reconnected at random, thus
introducing quenched disorder and creating shortcuts between
distant parts of the original lattice. Annealed small worlds
constitute a variation of the same model, introduced for systems
where propagation processes occur between sites
\cite{Manrubia,ann,sw3}. In this version of the small-world
model, propagation takes place between neighbour sites of the
ordered lattice but, with a certain probability, jumps to
randomly chosen sites may take place. Long-range jumps play the
role of the shortcuts of the quenched original version. By
analogy, these random jumps are associated with disorder in the
propagation process.

\begin{figure}
\resizebox{\columnwidth}{!}{\includegraphics*{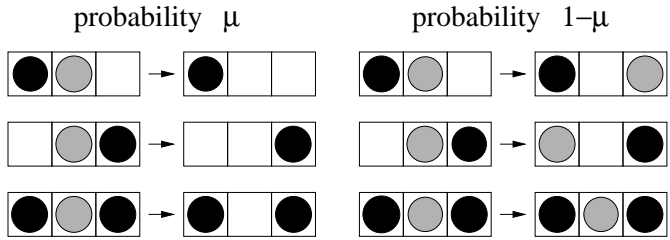}}
\caption{Possible events associated with the motion of a frustrated
particle. The moving particles is shown in gray. Left column:
long-range jump events. Right column: diffusion events.}
\label{fig1}
\end{figure}

In our model of segregation, a particle is chosen at random at
each time step. If it is not frustrated, it remains at its site.
If, on the other hand, the selected particle is frustrated,  it
is allowed to jump to a randomly selected empty site, anywhere in
the lattice, with probability $\mu$. With the complementary
probability, $1-\mu$, the motion is diffusive. If only one of the
nearest-neighbour sites is occupied, the particle moves to the
empty neighbour. If, otherwise, the two nearest-neighbour sites
are occupied, the exclusion principle forbids motion. The
possible instances are illustrated in Fig.~\ref{fig1}. As
discussed above, the parameter $\mu$ can be associated with the
level of disorder of the annealed small world. The duration of
each time step, $\Delta t=1/N$, is fixed so that, on the average,
each particle is chosen once per time unit.

We may assert {\it a priori} that the system has two
well-differentiated regimes, depending on how the density $\rho$
compares with $1/2$. For $\rho\le 1/2$, collective states where no
particle is frustrated exist. For finite $L$, moreover, the
population will necessarily reach one of such fully non-frustrated
states, from any initial condition, in a finite time $T$. From
then on, it will remain in a frozen configuration where all
particles are mutually separated by at least one empty site. On
the other hand, for $\rho > 1/2$, there will always be frustrated
particles, as there is not enough place to accommodate the
population in a fully non-frustrated state. As time elapses, the
number of frustrated particles will eventually fluctuate around a
well-defined stationary value.

Following this discussion, our analysis of the present model
focuses on the study of the evolution of the number of frustrated
particles. In the low-density regime, where the population is
able to accomplish full segregation, we record the average time
$\bar T$ needed to reach the fully non-frustrated state, starting
from a random distribution of particles. The average is performed
over different realizations of the initial state and of the
stochastic evolution. For high densities, on the other hand, we
analyze the asymptotic number of frustrated particles. For
convenience both in the numerical and in the analytical
calculations, instead of recording the number of frustrated
particles we follow the evolution of the number $P$ of particle
pairs, defined as the number of particles whose nearest-neighbour
sites to the right are occupied. The time average of the pair
density $p=P/N$ acts as an order parameter for the transition
between the two regimes. Our main parameters are the particle
density $\rho$ and the jump probability $\mu$.

\section{Preliminary numerical results} \label{num1}

We begin by exploring the scaling properties of our model, for
fixed $\rho$ and $\mu$, as a function of the system size $L$ --or,
equivalently, of the number of particles $N=\rho L$. It results
that, for sufficiently large systems, the evolution of the pair
density $p$ becomes size-independent when time is suitably scaled
by the step duration $\Delta t=1/N$. In further simulations, we
fix $L=5 \times 10^4$, for which this asymptotic independence of
size already holds.

\begin{figure}\begin{center}
\resizebox{\columnwidth}{!}{\includegraphics*{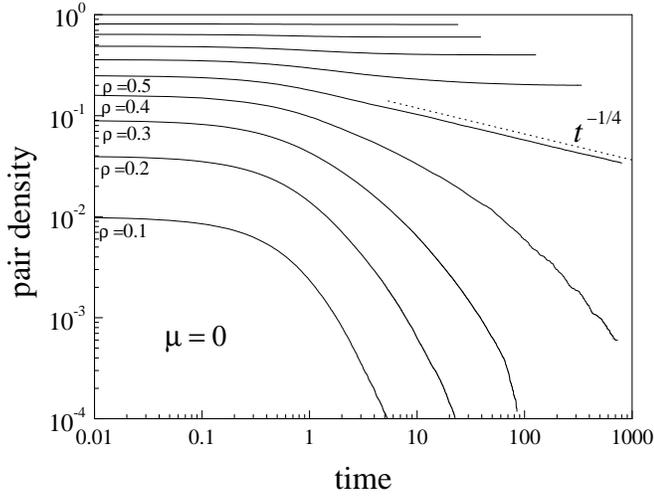}}\end{center}
\caption{Evolution of the pair density $p$ for
$\mu=0$ and several values of the particle density $\rho$. Curves
without labels correspond, from bottom to top, to $\rho=0.6$,
$0.7$, $0.8$, and $0.9$. The dotted line has slope $-1/4$.}
\label{nump0}
\end{figure}

\begin{figure}\begin{center}
\resizebox{\columnwidth}{!}{\includegraphics*{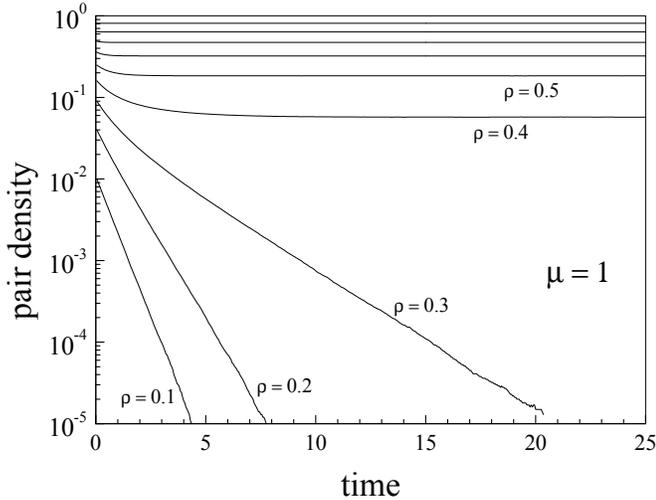}}\end{center}
\caption{Evolution of the pair density $p$ for $\mu=1$ and several
values of the particle density $\rho$. Curves without labels
correspond, from bottom to top, to $\rho=0.6$, $0.7$, $0.8$, and
$0.9$.} \label{nump1}
\end{figure}

Figures \ref{nump0} and \ref{nump1} show the evolution of the pair
density $p$ for several values of the particle density $\rho$ and
the two extreme values of small-world disorder, $\mu=0$ and $1$,
respectively. Each curve has been obtained by averaging over $100$
realizations. For $\mu=0$, as argued above, $p$ vanishes at finite
times if $\rho<1/2$ and reaches a slightly fluctuating stationary
level if $\rho>1/2$. The two regimes are separated by the case
$\rho=1/2$, where $p$ displays a power-law decay with an exponent
close to $-1/4$.

For $\mu=1$, rather unexpectedly, the transition between the two
regimes does not take place at $\rho=1/2$ but at a lower density
$\rho_c$, somewhere between $0.3$ and $0.4$. Even when we know
that, for $\rho_c <\rho <1/2$, fully non-frustrated states
certainly exist, the system is not able to find any of them and
the pair density approaches a stationary non-zero level. It must
be recalled that, as discussed above, a finite population will
always reach a fully non-frustrated state due to the effect of
fluctuations. For the system size of our simulations, however, it
is apparent from Fig. \ref{nump1} that the associated time grows
beyond our numerical reach as soon as $\rho$ becomes larger than
$\rho_c$.

\begin{figure}\begin{center}
\resizebox{\columnwidth}{!}{\includegraphics*{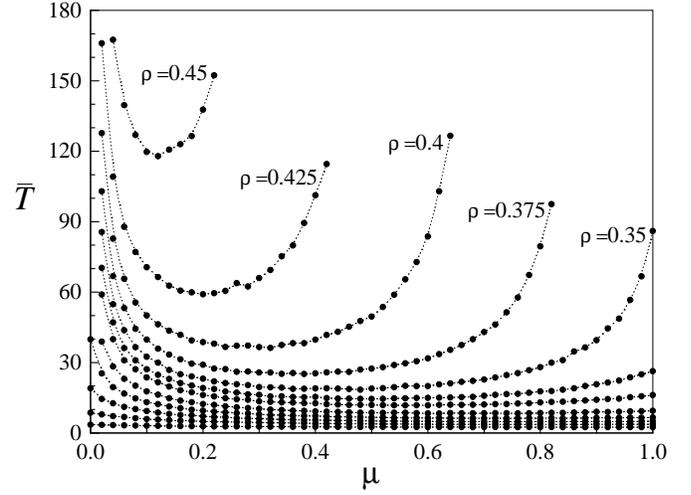}}\end{center}
\caption{Average time $\bar T$ needed to reach a fully
non-frustrated state  as a function of the disorder $\mu$, for
various values of the density $\rho$. Data without labels
correspond, from bottom to top, to $\rho=0.05$, $0.1$, $0.15$,
$0.2$, $0.25$, $0.3$, and $0.325$. For clarity, data have been
joined with dotted lines.} \label{times}
\end{figure}

To reveal this feature in more detail and, in particular, to
analyze the effect of the small-world disorder, we study the time
$T$ needed to reach a fully non-frustrated state. Figure
\ref{times} shows the average of $T$, taken over $500$ to $10^4$
numerical realizations, as a function of the disorder $\mu$ and
for several values of the density $\rho<1/2$. In agreement with
our previous results, we find that, for $\rho \gtrsim 0.35$, there
is a critical level of disorder $\mu_c$ at which the average time
$\bar T$ becomes very large. For $\rho \approx 0.35$, we have
$\mu_c\approx 1$, and the critical disorder decreases as the
particle density grows. It seems to approach zero as $\rho$
becomes closer to $1/2$.

For fixed $\rho$, the average time $\bar T$ grows not only as the
disorder increases towards $\mu_c$, but also as the disorder
decreases and approaches zero. This implies that, remarkably,
there is an intermediate value of $\mu$ for which $\bar T$ is
minimal. This value depends on the density, and decreases as
$\rho$ grows. At the same time, the minimum becomes sharper. For
small densities, the curve is so flat that the minimum cannot be
discerned, due to numerical fluctuations in the value of $\bar T$
as $\mu$ varies. For larger densities, on the other hand, the
minimum is very well defined. Thus, at least for sufficiently high
densities, the segregation process is fastest at an intermediate
value of the jump probability. Neither pure diffusion nor random
long-range jumps is more efficient, to drive the system to the
fully non-frustrated state, than this optimal combination of the
two processes.

Before proceeding further with numerical simulations, we turn the
attention to the analytical study of our system. Our analytical
approach makes it possible to explain --or, at least, to clarify
to a certain extent-- the numerical result obtained so far.
Moreover, it provides a clue to the steps to follow in further
simulations.

\section{Analytical results} \label{anal}

\subsection{Pair-antipair representation} \label{pap}

We have already introduced particle pairs as an alternative
representation of frustrated particles. In this section, we use an
extension of this representation as a convenient tool for the
analytical treatment of our system.

A particle pair has been assigned to each occupied site whose
nearest-neighbour site to the right is also occupied. The number
of pairs $P$ is thus given by the number of frustrated particles
with a neighbour particle to the right. The density of pairs is
$p=P/L$. In the same way, we assign an antipair to each empty
site whose nearest-neighbour site to the right is also empty. The
number of antipairs is denoted by $\bar P$, and their density is
$\bar p= \bar P/L$.

To complete the representation in terms of pair-like elements, we
need to introduce the number $W^-$ of occupied sites whose
nearest-neighbours to the right are empty, and the number $W^+$ of
empty sites with occupied sites to the right. However, due to the
presence of periodic boundary conditions, we necessarily have
$W^-=W^+$. It is therefore sufficient to consider their sum,
$W=W^-+W^+$, which stands for the total number of ``walls''
separating pair and antipair domains. The corresponding density
is $w=W/L$; we also introduce $q=w/2=W^-/L=W^+/L$.

\begin{figure}
\begin{center}
\resizebox{.7\columnwidth}{!}{\includegraphics*{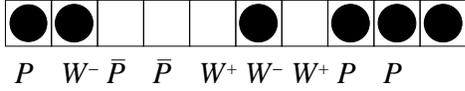}}
\end{center}
\caption{Pair-antipair representation of a given configuration on a
$10$-site domain.} \label{figpair}
\end{figure}

Figure \ref{figpair} illustrates the pair-antipair representation of
a given configuration of the system. Note that, due to the fact that
two contiguous elements of this representation share a common site,
not all configurations are possible. Specifically, only eight
combinations of contiguous elements can occur: $PP$, $PW^-$,
$W^-W^+$, $W^-\bar P$, $W^+P$, $W^+W^-$, $\bar P W^+$, and $\bar
P\bar P$. They correspond to the eight different three-site
neighbourhoods in the original representation. All of them are
represented in Fig. \ref{figpair}.

Expressing the total number of particles and of empty sites in
terms of the numbers $P$, $\bar P$, and $W$, it is possible to
verify that the densities $p$, $\bar p$, and $q$, and the particle
density $\rho$ are related according to
\begin{equation}
\begin{array}{rll}
p+q &= &\rho, \\
\bar p +q &= &1-\rho.
\end{array}
\end{equation}
These relations make it possible to write
\begin{equation} \label{q}
\begin{array}{rll}
q &= &\rho-p, \\
\bar p &= &1+p-2\rho.
\end{array}
\end{equation}
Consequently, the only independent quantities in the pair-antipair
representation are the particle density $\rho$ --which, we recall,
is one of our parameters-- and the pair density $p$. In the frame
of this representation, we analyze now the special cases $\mu=0$
and $\mu=1$.

\subsection{Pure diffusion: $\mu=0$} \label{mu0}

For $\mu=0$, frustrated particles can only move to
nearest-neighbour sites, in the case they are empty, and
long-range jumps are not possible. The motion is therefore purely
diffusive, with the constraint imposed by the exclusion principle.

\begin{figure}
\resizebox{\columnwidth}{!}{\includegraphics*{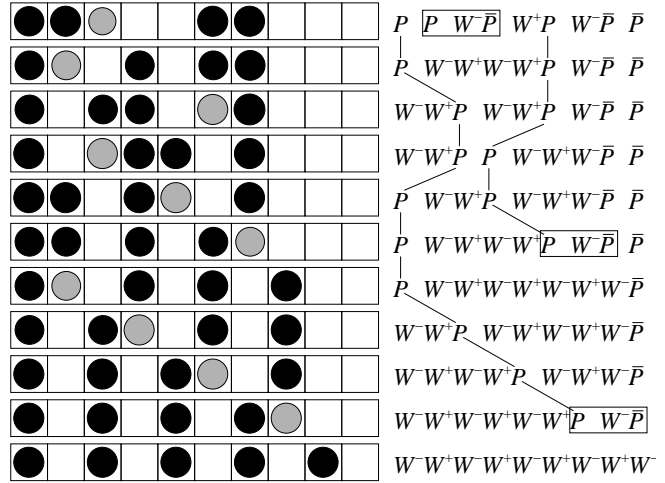}}
\vspace*{.5cm} \caption{A realization of the segregation process
of five particles, with $\mu=0$ (pure diffusion), within a
$10$-site spatial interval. Time runs downwards. At each step,
the moving particle is shown in gray. To the right, the
corresponding pair-antipair representation is given. Boxes
indicate pair-antipair annihilation events. Lines stand for pair
trajectories.} \label{pure}
\end{figure}

Figure \ref{pure} illustrates a particular realization, during
$11$ time steps, in a $10$-site portion of a larger system. Five
particles, all of which are initially frustrated, reach a fully
non-frustrated state. To the left of the plot, we show the
corresponding ($9$-site) pair-antipair representation. Pair
trajectories are shown with lines, and pair-antipair annihilation
events are indicated by boxes.

An exhaustive analysis of the possible configurations in small
neighbourhoods reveals that the are three elementary events able
to change the configuration. In the pair-antipair representation,
they are
$$
\begin{array}{c}
PW^-W^+ \leftrightarrow W^-W^+ P,\\
 PW^-\bar P \to W^-W^+W^- ,\\
\bar PW^+ P \to W^+W^-W^+.
\end{array}
$$
The first one can occur in both directions, and stands for the
diffusive motion of a pair $P$. Note that $P$ moves two sites to
the right or to the left at each step. Antipairs $\bar P$, on the
other hand, are immobile. The second and  third events stand for
pair-antipair annihilation. They are equivalent, in the sense that
they can be transformed into each other by spatial inversion.

Thus, in the limit of pure diffusion our population is analogous to
a reaction-diffusion system of two species, $P$ and $\bar P$. While
$P$ diffuses through short-range jumps, $\bar P$ remains immobile.
Moreover, if $P$ and $\bar P$ become sufficiently close to each
other, they may undergo binary annihilation:
$$
P+\bar P \to 0.
$$
Note that the possibility of this reaction was already implicit in
the second of Eqs. (\ref{q}), which states that the difference
$p-\bar p$ is a constant of motion (determined by the particle
density $\rho$).

The analogy with a two-species reaction-diffusion system makes it
possible to apply well-know results of dif\-fusion-controlled
bimolecular reactions \cite{anih}. Specifically, we know that if
the densities of pairs and antipairs are identical, $p=\bar p$,
the kinetics is anomalous. According to the second of Eqs.
(\ref{q}), such condition holds for $\rho=1/2$, where the only
stationary state for our system corresponds to a periodic array of
alternating occupied and empty sites. Under this condition,
fluctuations dominate the kinetics \cite{anih}. Pairs and
antipairs segregate into separated spatial domains and, from then
on, annihilation takes place at the domain boundaries only. Thus,
the kinetics is drastically slowed down and the pair density
decreases, for long times, as
\begin{equation} \label{1/4}
p \sim t^{-1/4}.
\end{equation}
Figure \ref{nump0} shows that numerical results are in good
agreement with this prediction.

For $p\neq \bar p$ the approach to stationarity is exponential. If
$p<\bar p$ ($\rho < 1/2$) all pairs are annihilated, while a
number of immobile antipairs subsists. Hence, the population is
able to reach a fully non-frustrated state. If, on the other hand
$p>\bar p $ ($\rho > 1/2$), moving pairs survive while all
antipairs disappear. The remaining density of pairs is $p^*=2\rho
- 1$, because $\bar p=0$. Asymptotically, the population is found
in a stationary but dynamical state with a non-vanishing fraction
of frustrated particles, as shown in Fig.~\ref{nump0}.

The analogy with binary annihilation makes it also possible to
predict, at least in qualitative terms, the time behaviour of the
pair density for $\mu \gtrsim 0$, as compared with the limit $\mu
=0$. As long as the disorder remains close to zero, pure diffusion
will dominate the transport of particles and, thus, pair and
antipair domains will form as a consequence of the mutual
annihilation of these two species. While the ensuing annihilation
events will mainly occur at the domain boundaries, rare long-range
jumps of frustrated particles will give origin to non-local
events where distant pairs and antipairs may disappear
simultaneously (see also Sect. \ref{mu1}). These non-local events
take place when a frustrated particle, jumping from a domain
occupied by pairs, lands at a antipair domain. Their contribution
to the variation of the pair density $p$ is proportional to the
product $p\bar p$ of the densities of pairs and antipairs, and to
the disorder $\mu$. For $\rho = 1/2$ ($p=\bar p$), this
contribution implies an algebraic decay $p \sim t^{-1}$; for
$\rho \neq 1/2$ the decay is exponential. In both cases, it
accelerates the decay of the pair density and, thus, the
asymptotic state is reached faster. Therefore, the evolution time
$T$ must be smaller for $\mu \gtrsim 0$ than for $\mu=0$, as it
was shown to be the case in our numerical simulations (Fig.
\ref{times}).

\subsection{Full disorder: $\mu=1$} \label{mu1}

If $\mu=1$, frustrated particles jump to randomly chosen empty
sites, and the small-world disorder is maximal. At each time
step, a particle can move a distance of the order of $L$.
Regarding the segregation process, however, these random jumps
represent an inefficient way of reaching a fully non-frustrated
state. As our preliminary numerical simulations have shown, this
results into a substantial increase of the times needed to
accomplish full segregation.

Since jumps take place to any empty site, it is possible to write
an equation for the evolution of the pair density $p$ by
evaluating the probability of the events which create or destroy
pairs in the limit of an infinitely large system. This
approximation requires assuming that the positions of pairs and
antipairs are not correlated.

\begin{figure}\begin{center}
\resizebox{.7\columnwidth}{!}{\includegraphics*{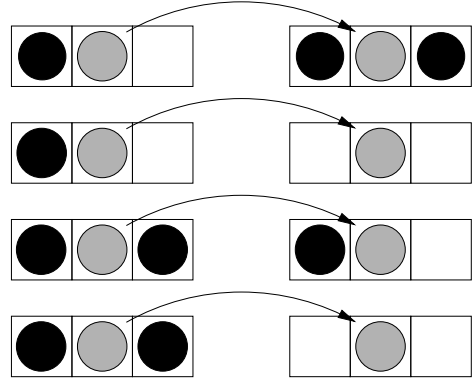}}\end{center}
\vspace*{.5cm} \caption{All the long-range jump events which
create or destroy particle pairs (up to spatial-inversion
transformations). The moving particle is shown in gray.}
\label{f3}
\end{figure}

Figure \ref{f3} shows --up to the transformation of spatial
inversion-- all the possible events with net creation or
destruction of pairs. The first line displays the only situation
where a pair --and the corresponding antipair-- can be created.
The moving particle starts from a site with one occupied
nearest-neighbour ($PW^-$) and lands at the empty site between
two occupied sites ($W^-W^+$). The final configuration at the
original site is $W^-\bar P$, and two pairs ($PP$) have appeared
at the landing neighbourhood. We represent this transition as
$$
PW^-|W^-W^+ \to W^-\bar P|PP,
$$
where to the left (right) of the arrow we indicate the initial
(final) configurations of both the starting and the landing
neighbourhoods. The (positive) contribution of this transition to
the average variation of the number of pairs $P$ during a time
step $\Delta t$ is proportional to the probability of finding the
initial neighbourhood configurations $PW^-$ and $W^-W^+$ when
choosing a particle at random. As for $PW^-$, the probability of
finding these two elements in contiguous positions --i.e.,
sharing a particle-- is $pq/\rho^2$. Equivalently, the
probability of finding $W^-W^+$ when choosing an empty site for
landing, is $q^2/(1-\rho)^2$. Thus, the contribution of the above
transition to the variation of $P$ is
\begin{equation} \label{p1}
\Delta P_1 = \frac{pq^3}{\rho^2 (1-\rho)^2}
\end{equation}
Spatial inversion transforms this transition into
$$W^+P|W^-W^+ \to \bar P W^+|PP.$$
By symmetry, its contribution to the variation of the number of
pairs is the same as above, so that $\Delta P_1$ will enter the
evolution equation for the pair density multiplied by two.

The three last lines in Fig. \ref{f3} stand for events where net
destruction of pairs --and  antipairs-- occurs. They correspond to
the transitions
$$
\begin{array}{lll}
PW^-|\bar P\bar P &\to &W^-\bar P | W^+W^- ,\\ \\
P P |W^-\bar P &\to &W^-W^+ |PW^-, \\ \\
P P|\bar P\bar P &\to &W^-W^+|W^+W^-, \\
\end{array}
$$
respectively. In the first two of these transitions only one pair is
destroyed, while in the third transition two pairs are
simultaneously destroyed. Following the same arguments as above, we
can calculate the respective (negative) contributions to the time
derivative of the pair density, as
\begin{equation} \label{p2}
\begin{array}{ll}
\Delta P_2 &= -p\bar p^2 q/\rho^2(1-\rho)^2, \\ \\
\Delta P_3 &=-p^2\bar p q/\rho^2(1-\rho)^2, \\ \\
\Delta P_4 &=-2 p^2\bar p^2/\rho^2(1-\rho)^2.
\end{array}
\end{equation}
The contributions $\Delta P_2$ and $\Delta P_3$ must be multiplied
by two, to take into account the transitions obtained by spatial
inversion. On the other hand, the transition giving origin to
$\Delta P_4$ is invariant under such transformation and,
therefore, there is no further contribution associated with it.

In summary, within our approximation, the average variation of
the number of pairs $P$ during a time step is given by $\Delta P
=2(\Delta P_1+\Delta P_2+\Delta P_3)+\Delta P_4$. The time
derivative of the pair density $p=P/L$ can be evaluated as $\dot p
= \Delta P/L\Delta t=\rho \Delta P$, where we have taken into
account that $\Delta t=1/N$. Using Eqs. (\ref{q}), (\ref{p1}),
and (\ref{p2}), we get
\begin{equation} \label{evolp}
\dot p = \frac{2p}{\rho (1-\rho)} (3\rho^2-3\rho p-\rho+p^2) .
\end{equation}
This equation has three fixed points, corresponding to stationary
values of the pair density. They are
\begin{equation} \label{p*}
p_0^*=0, \  \ \ \ \ \ p^*_\pm = \frac{3}{2}\rho\pm
\sqrt{\rho-\frac{3}{4}\rho^2}.
\end{equation}
Linear stability analysis shows that $p^*_+$  is unstable for any
value of the pair density $\rho$.  For $0<\rho<1/3$, $p^*_-$ is
negative and unstable, while $p_0^*$ is stable. For $\rho>1/3$,
these two fixed points interchange stability with each other, and
$p^*_->0$. Therefore, we have a transcritical bifurcation at
$\rho_c=1/3\approx 0.33$, in good agreement with our numerical
results. If the particle density is lower than $\rho_c$, the pair
density vanishes asymptotically, and the system is able to reach
a state where no frustrated particles subsist. For larger particle
densities, on the other hand, a fraction $p^*_->0$ of the
population remains always frustrated. It is remarkable that,
according to our prediction, an infinitely large population is
unable to find a fully non-frustrated state for $1/3<\rho<1/2$,
even when such states do exist within that density interval. This
feature makes clear that too much disorder in transport is
inefficient as for the achievement of full segregation.

\subsection{Summary of analytical results}

It is useful to briefly summarize our analytical conclusions, in
order to clarify to which extent our previous numerical results
have been explained and to define the aims of further simulations.
(i) For $\mu=0$, we have established the equivalence between our
system and the mutual annihilation of two diffusing species. Known
results for the latter make it possible to explain the observed
algebraic decay of the pair density, $p\sim t^{-1/4}$, for
$\rho=1/2$ (Fig. \ref{nump0}). We have also shown that, for
$\rho>1/2$, the asymptotic pair density should be $p^*=2\rho-1$.
(ii) For $\mu \gtrsim 0$, it was possible to show qualitatively
that the evolution to full segregation is faster than for $\mu=0$,
as already found in the simulations (Fig. \ref{times}). (iii) For
$\mu=1$, we have derived an evolution equation for the pair
density $p$. This equation makes it possible to predict, in
particular, the asymptotic values of $p$ as a function of $\rho$,
from Eq. (\ref{p*}). Also, it predicts a transcritical bifurcation
between the regimes of low and high particle densities at
$\rho_c=1/3$. Combining conclusions (ii) and (iii), we can give
analytical support to the observation of a minimum in the average
time $\bar T$ as a function of the disorder $\mu$ (Fig.
\ref{times}). In fact, while $\bar T$ decreases for small $\mu$,
it diverges at a finite disorder level, $\mu_c<1$.

\section{Further numerical results and phase diagram} \label{num2}

\begin{figure}\begin{center}
\resizebox{\columnwidth}{!}{\includegraphics*{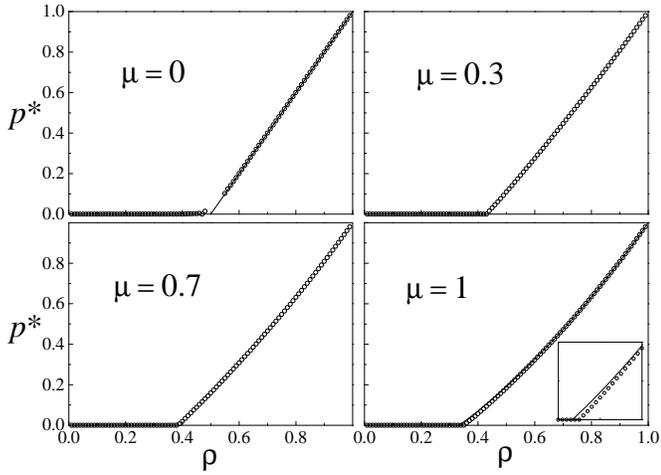}}\end{center}
\caption{The stationary pair density $p^*$ as a function of the
particle density $\rho$, for four values of the small-wolrd
disorder $\mu$. For $\mu=0$ and $1$, full lines show the
analytical predictions (see Sects. \ref{mu0} and \ref{mu1}). For
$\mu=1$, the inset shows a close-up of the main plot for
$0.3<\rho<0.5$.} \label{stat}
\end{figure}

Following our analytical results, it is first pertinent to study
in more detail the asymptotic pair density $p^*$ for different
values of the small-world disorder $\mu$ and of the particle
density $\rho$. Figure \ref{stat} shows numerical measurements of
$p^*$ as a function of $\rho$, for four values of $\mu$. For
$\mu=0$, results are in excellent agreement with the analytical
prediction, $p^*=2\rho-1$, shown as a line. As $\mu$ grows, the
dependence of $p^*$ on $\rho$ becomes increasingly non-linear and,
at the same time, the critical density $\rho_c$ shifts to the
left. For $\mu=1$, the numerical value of the critical density is
slightly above the prediction $\rho_c=1/3$. Moreover, while
analytical and numerical results for $p^*$ as a function of
$\rho$ are in good agreement, the analytical prediction is
systematically larger. As shown in the inset of the fourth panel
of Fig. \ref{stat}, this discrepancy is more important close to
the critical point.

\begin{figure}\begin{center}
\resizebox{\columnwidth}{!}{\includegraphics*{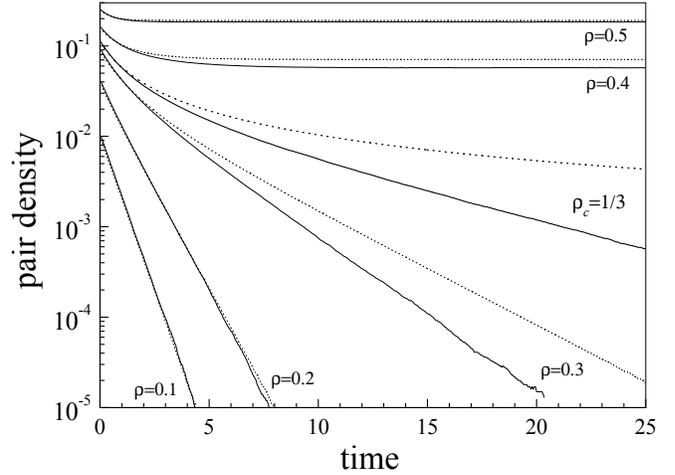}}\end{center}
\caption{The evolution of the pair density $p$ for various values
of the particle density $\rho$, with $\mu=1$. Full lines
correspond to simulation results, and dotted lines stand for the
numerical solution of Eq. (\ref{evolp}). } \label{comp}
\end{figure}

As may be expected, the same effect is present in the comparison
between the evolution of the pair density for $\mu=1$ as obtained
from simulations, and the solution of the evolution equation
(\ref{evolp}). Figure \ref{comp} shows, as full lines, simulation
results for various values of the particle density $\rho$, in the
same conditions as in Fig. \ref{nump1}. Dotted lines stand the
corresponding solutions of Eq. (\ref{evolp}). The latter have been
obtained through a fourth-order Runge-Kutta integration of the
equation with the initial condition $p(0)=\rho^2$, which
corresponds to a random distribution of particles. The agreement
between the two results is very good for small densities, well
below the critical value $\rho_c=1/3$. As $\rho$ approaches
$\rho_c$, however, a noticeable discrepancy arises in the
long-time evolution of the pair density. On the other hand, the
initial evolution obtained from simulations is well described by
the analytical prediction. The long-time discrepancy seems to
reach its largest relative value precisely at the critical
density $\rho_c$. From then on, its effect decreases steadily,
though a difference in the asymptotic values $p^*$ persist, as
already shown in Fig. \ref{stat}.

We identify two possible causes for the discrepancy between
simulation and analytical results for $\mu=1$. First, simulations
are necessarily affected by finite-size effects. In our system,
they convey the possibility that, due to fluctuations, the
population reaches a fully non-frustrated state faster than
predicted by the average-like arguments of the analytical
formulation, which hold for an infinitely large system. We have
already stated that, indeed, a finite system should always be
able to reach a fully non-frustrated state in a finite time for
any particle density lower than $1/2$. Such ``early'' equilibria
may explain the fact that simulation results for the pair density
are systematically below the analytical prediction. As expected,
the effect is relatively larger close to the critical point.

The second source of discrepancy lies on the uncontrolled
assumption made to formulate Eq. (\ref{evolp}) that the positions
of pair-like objects are not correlated. While long-range
correlations are surely negligible for the random initial
condition, the evolution --which systematically tends to destroy
frustrated configurations and to stabilize non-frustrated states--
is expected to create such correlations. This is illustrated in
Fig. \ref{pure} (for $\mu=0$), where the creation of long-range
structures such as $\cdots W^+W^-W^+W^- \cdots$ is apparent. While
the effect of reaching ``early'' equilibria, discussed above,
disappears for $\rho>1/2$, the effect of creation of correlations
between pair-like objects is present in the whole domain of
particle densities.

\begin{figure}\begin{center}
\resizebox{\columnwidth}{!}{\includegraphics*{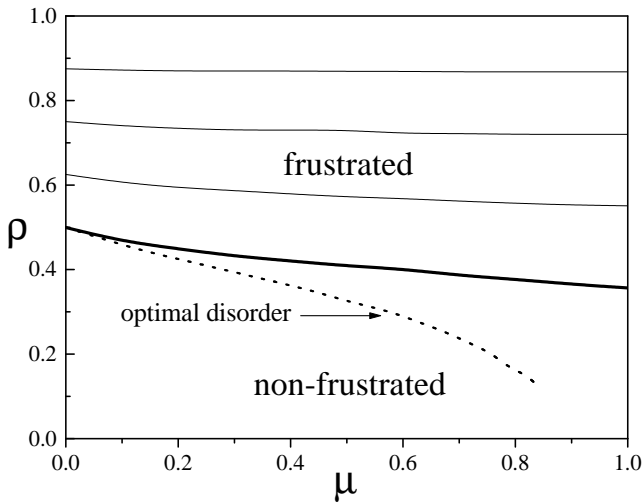}}\end{center}
\caption{Phase diagram in parameter space. The bold line
separates the region where a fully non-frustrated state is
reached from that where frustrated particles subsist for
asymptotically long times. The dotted line indicates the values of
($\mu,\rho$) for which the fully non-frustrated state is reached
fastest. Lines in the frustrated phase are  level curves of
constant stationary pair density $p^*$.} \label{fases}
\end{figure}

Our main results for segregation in annealed small worlds are
summarized in the phase diagram of Fig. \ref{fases}. The
parameter space ($\mu,\rho$), spanned by the small-world disorder
and the particle density, is divided into two main zones. For low
densities, we have the phase where the population is able to reach
a fully non-frustrated state while, for large densities, some
degree of frustration persists even at asymptotically long times.
Within the former, we find the curve over which the average time
needed to reach the fully non-frustrated state is minimal. For a
given particle density, it gives the optimal small-world disorder
to achieve segregation. In the frustrated phase, Fig. \ref{fases}
shows curves corresponding to constant values for the density of
pairs persisting at long times.

\section{Conclusion}

In this article, we have presented and studied a model for
microscopic segregation in a homogeneous population of particles
moving on a one-dimensional annealed small-world network. Motion
is thus a mixture of short-range diffusion and long-range jumps,
picking (frustrated) particles with occupied neighbours and
transporting them to empty sites. Small-world disorder is here
identified with the probability of long-range jumps. While no
macroscopic separation of phases takes place, for low densities
the population is driven toward a (fully non-frustrated) spatial
distribution where particles are mutually separated by at least
one empty site.

From numerical simulations of the system we have obtained, on the
one hand, the time needed to achieve full segregation in the
regime of low densities. On the other hand, for high densities,
we have analyzed the asymptotic number of persisting frustrated
particles. In the parameter space spanned by the small-world
disorder and the particle density, we have determined the
boundary between the two regimes, which interrelates the two
parameters in a non-trivial way. These numerical results have
been partially explained by means of an analytical approximation,
based on a representation of the particle distribution through
pair-like elements. In this representation, the motion of
individual particles is associated with events of pair motion or
pair-antipair annihilation. The approximation is useful for
vanishing small-world disorder, where the segregation process
becomes equivalent to the binary annihilation of two diffusing
species. In this limit, our approximation makes it possible to
predict the anomalous time decay of the number of frustrated
pairs at the boundary between the regimes of low and high
densities. In the opposite limit of maximal small-world disorder,
where particle motion occurs purely through long-range jumps, the
approximation predicts a transcritical bifurcation at the
boundary. Numerical measurements of the asymptotic density of
frustrated pairs are in good agreement with the analytical
prediction. Also, the short-time evolution of the pair density is
reasonably well approximated by the analytical results.

Our most remarkable results concern the regime of low densities.
As stated above, in this regime, the average time needed to reach
a fully non-frustrated state is a meaningful overall quantity
characterizing the dynamics. We have found, first, that this time
attains a minimum at an intermediate value of the small-world
disorder. In other words, there is an optimal combination of
diffusion and long-range jumps that makes the segregation process
fastest. Second, both numerical and analytical evidence show
that, in the limit of an infinitely large system, there is a
critical value of the small-world disorder above which the time
needed to reach the fully non-frustrated state diverges. Too many
long-range jumps, thus, make the achievement of full segregation
impossible. The critical value of the small-world disorder
determines the boundary between the regimes of low and high
densities.

Besides the relevance of these conclusions as for the dynamics of
segregation itself, they are particularly interesting from the
viewpoint of small-world based processes. In fact, it is well
known that the threshold of critical behaviour in a large class
of systems with underlying small-world structure shifts toward
zero small-world disorder as the system size $N$ grows
\cite{barrat}. This is a direct consequence of the fact that the
cross-over from the ordered-like regime to the random-like regime
in the geometrical properties of small-world networks occurs at a
disorder level of order $N^{-1}$ \cite{mouk,mouk1}. Only a handful
of small-world based processes have been reported where critical
behaviour or non-monotonic dependence on the disorder persist at
finite disorder levels as $N$ is increased
\cite{sw2,sw3,sanchez}. Our system is a novel instance of this
rare category, combining both non-monotonic dependence and
critical behaviour.

Several generalizations of the present model would be worth
considering. First, the same dynamical rules can be implemented
in two- or higher-dimensional lattices. In more dimensions, an
additional degree of freedom is given by the various ways in
which the neighbourhood of a site can be defined. In turn, this
determines several possibilities as for the definition of
frustrated and non-frustrated states. Another important
generalization has to do with the inclusion of noise. In the
present version of the model, diffusion events are permitted only
when they drive a particle from a frustrated state to a
non-frustrated state. In this sense, diffusion in our model is
unidirectional toward segregation. Noise can be implemented by
allowing a certain degree of reversibility for diffusion events,
such that non-frustrated particles might occasionally be lead to
frustration. This additional mechanism would be particularly
relevant for small disorder levels, when motion is dominated by
diffusion. In the pair-antipair representation, reversible
diffusion events would be represented by spontaneous creation of
a pair and an antipair, $0\to P+\bar P$, corresponding to the
transition $W^-W^+W^- \to PW^-\bar P$ or to its spatially inverted
form $W^+W^-W^+ \to \bar P W^+ P$. Finally, we recall from the
introduction that segregation can be thought of as a process aimed
at distributing a population in order to optimize the exploitation
of space. Our model may be extended to drive the population to a
state where the inter-particle distance is maximal, under the
constraint imposed by the particle density.


\begin{thebibliography}{}

\bibitem{trans} D. L. Goodstein, {\it States of Matter} (Dover, New York, 1985).

\bibitem{anih}  V. N. Kuzovkov, E. A. Kotomin, Rep. Prog.
Phys. {\bf 51}, 1479 (1988).

\bibitem{morphog} J. D. Murray, {\it Mathematical Biology} (Springer, Berlin,
1993).

\bibitem{evol} P. W. Skelton, {\it Evolution. A Biological and Paleontological
Approach} (Addison-Wesley, Singapore, 1996).

\bibitem{cult}  R. Dunbar, Ch. Knight, C. Power, eds., {\it The Evolution
of Culture. An Interdisciplinary View} (Rutgers University Press,
New Brunswick, 1999).

\bibitem{wattstr} D. J. Watts, S. H. Strogatz,  Nature {\bf
393}, 440 (1998).

\bibitem{sw1}  D.J. Watts, {\it Small Worlds} (Princeton University Press, Princeton,
1999).

\bibitem{sw2} M. Kuperman, G. Abramson, Phys. Rev. Lett. {\bf 86}, 2909
(2001).

\bibitem{sw3} D. H. Zanette, Phys. Rev. E {\bf 65}, 041908 (2002).

\bibitem{Manrubia} S. C. Manrubia, J. Delgado,  B. Luque, Europhys. Lett. {\bf 53}, 693
(2001).

\bibitem{ann} J. Lahtinen, J. Kert\'esz,  K. Kaski, Physica A {\bf
311}, 571 (2002).

\bibitem{barrat} A. Barrat, M. Weigt, Eur. Phys. J. B {\bf 13},
547 (2000).

\bibitem{mouk} C. F. Moukarzel, Phys. Rev. E {\bf 60}, R6263
(1999).

\bibitem{mouk1} M. Argollo de Menezes, C. F. Moukarzel, T. F.
J. Penna, Europhys. Lett. {\bf 50}, 574 (2000).

\bibitem{sanchez} A. D. S\'anchez, J. D. L\'opez, M. A.
Rodr\'{\i}guez, Phys. Rev. Lett. {\bf 88}, 048701 (2002).

 \end{thebibliography}
\end{document}